\documentstyle[12pt]{article}

\renewcommand{\theequation}{\thesection.\arabic{equation}}
\begin{document}

\begin{titlepage}
 \renewcommand{\thefootnote}{\fnsymbol{footnote}}
    \begin{normalsize}
     \begin{flushright}
                 UT-676\\
                 April 1994
     \end{flushright}
    \end{normalsize}
    \begin{Large}
       \vspace{1cm}
       \begin{center}
         {\bf  Long-Distance Universality of
                Laughlin State and  Calogero-Sutherland Model
         } \\
       \end{center}
    \end{Large}

  \vspace{10mm}
\begin{center}
\begin{Large}
           Satoshi Iso\footnote
           {E-mail address: iso@danjuro.phys.s.u-tokyo.ac.jp} \\
\end{Large}
         \vspace{2cm}
         {\it Department of Physics, University of Tokyo,} \\
               {\it Bunkyo-ku, Tokyo 113, Japan}\\

\vspace{15mm}

\end{center}
\begin{abstract}
\noindent
We study the universal long-distance behaviour of
the Laughlin state for the fractional quantum Hall effect and
the ground state of the Calogero-Sutherland model (one dimensional $1/r^2$
interaction
model). In particular, it is shown that
these two wave functions  coincide exactly when Laughlin state is confined
in a narrow cylinder geometry. The seeming difference of dimensionality
is merely a difference of representation of wave functions.
We also give a recipe to interpret operators acting on states in the lowest
Landau
level in terms of the usual one dimensional Fermion operators, which is
important for
extracting the Tomonaga-Luttinger liquid behaviour of the edge states.
\end{abstract}

\end{titlepage}
\vfil\eject

\newpage
\section{Introduction}
\setcounter{equation}{0}
Many similarities have been pointed out between the quantum Hall
effect \cite{laughlin} and the one dimensional (1D) solvable model with $1/r^2$
interaction (Calogero-Sutherland model \cite{calogero,sutherland}).
(1) Both of the Laughlin trial wave function of the fractional quantum Hall
effect
\begin{equation}
\sum\limits_{i<j} {\left( {z_i-z_j} \right)}^m
\exp \left( {-B\sum {|z_i|^2/ 4}} \right)
\end{equation}
and the ground state of the Calogero model
\begin{equation}
\sum\limits_{i<j} {\left( {x_i-x_j} \right)}^m
\exp \left( {-\omega \sum {x_i^2/ 2}} \right)
\end{equation}
are Jastrow
type and zero's of these wave functions are located
where positions of two particles
become to coincide. (2) Excited states are constructed by multiplying
polynomials to the ground state and they add an extra zero to the wave
functions. (3) In both models, when two particle are interchanged, phase $m
\pi$ appears instead of the usual Fermion phase $\pi$ and therefore these
Fermions are sometimes called hyper-Fermion.\footnote{In Calogero-Sutherland
model the two-body phase shift $m \pi$ is understood as the phase accompanied
by interchange of particles.}
This is responsible for the
fractional statistics  \cite{haldane}
and the fractional charge of the excited states which are
 characteristic properties of the quantum Hall effect
 and the Calogero-Sutherland model.
(4) In both models, hierarchy extension exists. This is also
originated in the above hyper Fermionic structure. In the quantum Hall effect
hierarchical states are realized dynamically but in Calogero-Sutherland model
they are realized by introducing flavor degrees of freedom from the beginning
and adjusting their couplings \cite{kawakami}.
More similarities such as an infinite dimensional algebra, $W_{\infty}$ algebra
\cite{w,hikami}, have been discussed.

In spite of these similarities these two systems look apparently
different. Quantum Hall effect is a many body problem of two
dimensional (2D) Fermions in a strong magnetic field while
Calogero-Sutherland model is a 1D Luttinger liquid
and the dimension of their configuration space is different.
Hence it is widely believed that the similarities can be seen only in the
one dimensional part of the quantum Hall state, edge state.
It is discussed by Wen \cite{wen} that the edge state of the quantum Hall
effect
is described by the chiral Luttinger liquid and its exponent is
equal to that of the chiral-constrained Calogero-Sutherland model
\cite{kawakami}.
This is partly confirmed by numerical evaluation
of the occupation numbers for small electrons \cite{mitra}, and by an analytic
calculation of the density \cite{wen2}.
\par
The purpose of this paper is to search for  deeper
relation between the quantum Hall effect and the 1D Luttinger liquid
(Calogero-Sutherland model). The dimensionality which makes
these systems look different can be shown to be nothing more than
difference in representation of wave functions.
In the system of the quantum Hall effect, electrons are subject to
strong magnetic field and are constrained in the lowest Landau
level. Hence the 2D space is interpreted as a phase
space and the guiding center coordinates of the cyclotron motion
$(X,Y)$ become canonically conjugate each other. By this interpretation
the
lowest Landau level wave functions are interpreted as the holomorphic
representation
of a 1D Fermions system represented on a 2D
phase space. In a previous paper \cite{azumaiso}, we have studied the 1D
representation of the Laughlin wave functions on various geometries
and shown that in certain limits the Laughlin states become equal to
the ground state of Calogero-Sutherland model.  Looking reversely, any
1D wave function can be seen as a wave function of the
lowest Landau level through the holomorphic representation.
In this representation, Calogero-Sutherland ground state has a similar form
 to the Laughlin state and  the only difference is that zero's of the
Calogero-Sutherland wave function
are distributed with an extent of the magnetic length
around where positions of two particles coincide
 while zero's of the Laughlin state for some particle are
located exactly on the other particles.
 This suggests that
Calogero-Sutherland model is in the same universality class as the Laughlin
state.

The correspondence of a 2D Fermion system in the lowest Landau level
and a 1D Fermion system is not restricted to wave functions.
2D density operator or 2D Fermion
operator projected on the lowest Landau level can be interpreted in terms
of 1D Fermion operators and vice versa.
By these correspondences we can make a recipe to express
any 2D correlator  in terms of 1D ones and any 1D correlator in
terms of 2D ones.
Especially 2D correlators along boundaries of a
quantum Hall droplet can be expressed with its 1D
correlator. This makes it possible to extract the chiral Luttinger
liquid behaviour of the quantum Hall effect.

The contents of the paper is the following.
In section 2 we show how 2D wave functions in the lowest Landau
level can be written as wave functions of the usual 1D
system. In particular, we study  Laughlin state in a narrow channel
and show that in this geometry Laughlin state becomes equal
to the ground state of Calogero-Sutherland model.
In section 3 we give a recipe to express two(one)-dimensional
operators in terms of one(two)-dimensional operators.
In section 4 we interpret the Calogero-Sutherland model in
terms of a 2D lowest Landau level system through the
holomorphic representation. In this representation the holomorphic
part of the ground state of the Calogero-Sutherland model
has a similar structure to the Laughlin state.  This suggests that the
Calogero-Sutherland model is in the same universality class
as the quantum Hall effect.  Section 5 is devoted to discussion.
\section{1D Representation of Lowest Landau Level System}
In this section we first fix the notation of kinematics in the lowest
Landau level.
For a planar electron in a magnetic field Hamiltonian is given by
\begin{equation}
H_0=\sum_{i=1,2} {(\Pi_i)^2 \over 2m}, \ \ \ \Pi_i=p_i-A_i, \ \ \ i=1,2.
\end{equation}
We assume that the constant magnetic field
is in the negative $z$ direction. By defining an annihilation operator
\begin{equation}
a=(\Pi_x-i \Pi_y)/\sqrt{2B \hbar}, \ \ \ [a,a^{\dagger}]=1,
\end{equation}
$H_0$ is written as
$H_0=\hbar \omega_c (a^{\dagger} a+1/2)$, where $\omega_c$ is the cyclotron
frequency; $\omega_c=B/m$. Heisenberg  equations of motion for $\Pi$
\begin{equation}
\dot{\Pi}_x={i \over \hbar} [H_0,\Pi_x]=-\omega_c \Pi_y, \ \
\dot{\Pi}_y={i \over \hbar} [H_0,\Pi_y]=\omega_c \Pi_x
\end{equation}
show that $(\Pi_x,\Pi_y)$ rotate with frequency $\omega_c$
and therefore
represent cyclotron motion in a magnetic field.
Spectrum of $H_0$ is  quantized as Landau levels. States in the lowest
Landau level (LLL) satisfy the LLL condition $a\phi_0=0$ and
$a^{\dagger}$ creates states in higher Landau levels.
The guiding center coordinates
of the cyclotron motion are defined by \cite{kubo}
\begin{equation}
X=x-{\Pi_y \over B}, \ \ \ Y=y+{\Pi_x \over B},\ \ \
[X,Y]={i \hbar \over B}.
\end{equation}
They commute with $a$ and $a^{\dagger}$ and therefore with $H_0$.
These coordinates describe degeneracy in each Landau level.
These four variables $(\Pi_x, \Pi_y, X, Y)$ are  more convenient
phase space variables than $(p_x, p_y, x, y)$
in a constant magnetic field.
\par
In a strong magnetic field, all electrons are confined in
LLL whose one-particle wave functions satisfy the
LLL constraint $a \phi_0=0$.  If we constrain the Hilbert space onto
the LLL, two of four phase space variables, $(\Pi_x, \Pi_y)$ are
frozen and remaining degrees of freedom are
the guiding center coordinates
 $(X,Y)$.  Therefore, effective degrees of freedom in the LLL
are reduced to the half of the total degrees of freedom
 and the two-dimensional
$(X,Y)$-coordinate space can be seen as a phase space of 1D system
\cite{iso}. This makes it possible to interpret the lowest Landau
level wave functions as one-dimensional system.
That is, when $X$ is diagonalized $X|s \rangle =s|s \rangle$,  Y is interpreted
as its dual momentum $Y=p_s/B$.
Eigenstate of $X$ in the LLL, $|s \rangle$, is uniquely determined
 and forms a complete  basis in the LLL;
\begin{equation}
\int |s \rangle \langle s| ds=1.
\end{equation}
Since any LLL wave function can be written in terms of
$|s \rangle$, wave functions and Hamiltonian in LLL can be interpreted as
those of 1D system whose coordinate is $s$.
\par
In the paper \cite{azumaiso}, 1D representation of Laughlin state
on various geometries are studied and they are shown to be closely
related with various $1/r^2$ interaction models. In this paper
we mainly concern the Laughlin state on a cylinder geometry.
In the Landau gauge ${\bf A}=(By, 0)$, LLL wave functions are written
as $\Psi(\bar{z}) e^{-By^2/2 \hbar}$.
Here $z=\sqrt{B/2\hbar} (x+iy)$.
We impose a periodic boundary condition for $x$ with period $L_x$.
Then an anti-holomorphic part of a LLL wave function can be written as
a linear combination of $\exp[2\pi i n   (x-iy)/L_x]=\omega^n$
where $\omega \equiv \exp[2\pi i   (x-iy)/ L_x] .$
The wave function $\omega^n$ has a momentum $2\pi\hbar n/ L_x$
and localized in $y$ direction around $y=2\pi\hbar n/B L_x$ with a width of
magnetic length $l=\sqrt{\hbar/B}$;
\begin{equation}
\omega ^n\exp \left( {-By^2/ 2\hbar } \right)\propto
 \exp \left( {2\pi inx/ L_x} \right)
\exp \left( {-B\left( {y-2\pi \hbar n/ BL_x} \right)^2/ 2\hbar }
 \right).
\end{equation}
  Filling factor $\nu=1$ state is given by
\begin{equation}
\prod_{i<j} (\omega_i-\omega_j) e^{-B \sum_i y_i^2/2 \hbar}.
\end{equation}
Since $\prod_{i<j} (\omega_i-\omega_j)$ is a Slater determinant
of $(1,\omega, ...,\omega^{N-1})$, there are two boundaries at $y=0$
and $y=2\pi (N-1)/BL_x$. These boundaries can be interpreted
as  Fermi surfaces of one dimensional Fermions.
Laughlin state with filling factor $\nu=1/m$ can be constructed as
\begin{equation}
\prod_{i<j} (\omega_i-\omega_j)^m  e^{-B \sum_i y_i^2/2 \hbar}.
\label{laugh2}
\end{equation}
  Filled region is expanded by $m$-times and
 boundaries are located at $y=0$ and $y=2\pi m(N-1)/BL_x$.
Low energy gapless excitations are deformation of the boundaries
(edge excitations). In this sense boundaries of Laughlin state
are also interpreted as Fermi surfaces of one dimensional Fermion
system but there is one big difference. While usual Fermi surfaces are
boundaries of completely filled regions and unfilled regions,
filled regions  in fractional quantum Hall droplet are not
completely  but fractionally filled. Nonetheless due to
imcompressibility of fractional quantum Hall state boundaries of
FQHE can be again interpreted as Fermi surfaces.
In the last section we propose a new type of one dimensional
Luttinger liquid which we should call fractional Tomonaga
Luttinger liquid.
\par
Now let's obtain the 1D representation of Laughlin state (\ref{laugh2}).
In Landau gauge, an eigenstate of $X=x+i \hbar \partial_y /B$ in LLL is
given by
\begin{equation}
\langle z \bar{z}|s \rangle =
\left( {B \over \pi \hbar}  \right)^{1/4}
\exp(-Bs^2/2 \hbar) \exp(\sqrt{2B/\hbar} s\bar{z}-\bar{z}^2)
\exp(-By^2/2\hbar).
\label{1Drep}
\end{equation}
By using this eigenstate
 1D representation of a LLL wave function
$
\langle z \bar{z}|\Psi \rangle=\Psi(\bar{z}) e^{-By^2/2\hbar}
$
can be obtained by
\begin{eqnarray}
\langle s|\Psi \rangle
&=& \int \langle s|z \bar{z} \rangle \langle z \bar{z}|\Psi \rangle
{d^2 z \over \pi} \nonumber \\
&=&
\left( {B \over \pi \hbar}  \right) ^{1/4} \int
e^{-Bs^2/2\hbar} e^{\sqrt{2B/\hbar} s z - z^2 /2}e^{-|z|^2}
e^{\bar{z}^2/2} \ \Psi(\bar{z})
{d^2 z \over \pi} \nonumber \\
&=&
\left( {B \over \pi\hbar}  \right)^{1/4}
e^{-Bs^2/2\hbar} e^{-(\hbar / 4B)  (\partial / \partial s)^2}
e^{Bs^2/\hbar} \ \Psi(\sqrt{2B/\hbar} \ s)  \nonumber \\
&=&
\left( {B \over \pi\hbar}  \right)^{1/4}
e^{-Bs^2/2\hbar} e^{-(\hbar / 4B)  (\partial / \partial s)^2}
e^{Bs^2/\hbar} \ 2^{s \partial_s} \Psi(\sqrt{B / 2\hbar} \ s)
\nonumber \\
&=&
\left( {B \over \pi\hbar}  \right)^{1/4} g_0 \
e^{-(\hbar / 2B) (\partial / \partial s)^2}
\Psi(\sqrt{B /2\hbar} \ s) ,
\label{cylinder1}
\end{eqnarray}
where
$g_0=\sum_{n=0}^{\infty} (-1/4)^n (2n)!/(n!)^2=1/\sqrt{2}$.
In the last equality we used the relation proved in appendix.
Extending it to a many-particle case,
 1D representation of Laughlin state on cylinder
is given by
\begin{equation}
\langle s_1 \cdot \cdot \cdot s_N|\Psi \rangle =
e^{-{\hbar \over 2B} \sum_i \left( {\partial \over \partial s_i} \right)^2}
\prod_{i<j} (e^{i2\pi s_i/L_x} -e^{i2\pi s_j/L_x} )^m.
\end{equation}
Here we neglected an overall constant.
In $B \rightarrow \infty$ limit, this wave function reduces to the
ground state of Sutherland model \cite{yoshioka};
\begin{equation}
\prod_{i<j} (e^{i2\pi s_i/L_x} -e^{i2\pi s_j/L_x} )^m.
\label{Sutherland}
\end{equation}
For $m=1$, $\sum_i (\partial_i)^2$ becomes a constant and this
is exact independent of large $B$ limit.
The Hamiltonian of Sutherland model is given by
\begin{equation}
{\cal H}=\sum\limits_{i=1}^N {p_i^2}+
g\left( {{\pi  \over {L_x}}} \right)^2\sum\limits_{i<j}
{{1 \over {\sin ^2\left( {\pi \left( {s_i-s_j} \right)/ L_x} \right)}}}
\end{equation}
where $g=2(m^2-m)$. This model is integrable and has $N$ integrals
of motion.
\par
A sufficient condition that the Laughlin state on a cylinder coincides
with the ground state of Sutherland model is
\begin{equation}
{\hbar  \over B}\left( {{{2\pi Nm} \over {L_x}}} \right)^2 \ll 1.
\label{narrow}
\end{equation}
Since the width of the cylinder Laughlin state is
$\delta y =2 \pi N m \hbar/B L_x$, this condition can be rephrased
that the droplet must be very narrow
$\delta y \ll l=\sqrt{\hbar/B}$. This condition can be interpreted differently
as follows. When the operator
$\exp(-{\hbar \over 2B} \sum \left( {\partial \over \partial s} \right)^2)$
acts on a polynomial $(s_i-s_j)^m$ it becomes
\begin{equation}
\approx \left( {s_i-s_j} \right)^m+
c_1{\hbar  \over B}\left( {s_i-s_j} \right)^{m-2}
+c_2\left( {{\hbar  \over B}} \right)^2\left( {s_i-s_j} \right)^{m-4}+...
\end{equation}
and if $(s_i-s_j) \gg l=\sqrt{\hbar /B}$, the terms with a smaller power of
$(s_i-s_j)$ than $m$ can be neglected.
The above narrow channel condition (\ref{narrow}) is equivalent to the
condition
that the mean 1D distance  $n=L_x/N$ is larger than the magnetic length
\begin{equation}
n={L_x \over N} \gg l=\sqrt{{\hbar \over B}}
\end{equation}
and
the exponential operator can be neglected for a dilute limit in 1D.
\par
Here we have seen that the narrow channel condition is sufficient
for equivalence of the Laughlin state and the ground state of
Sutherland model. In section 4 we come back to this point again.
\par
\section{Correlation Functions}
In this section we give a recipe to relate 2D correlators in the LLL
with usual 1D correlators of the Luttinger liquid. As we showed in the previous
section, Fock space of the LLL and that of the 1D system are equivalent.
Then any operator acting
on the Fock space of LLL can be interpreted as an operator acting
on a 1D system and the inverse is also true.
\par
First let us consider the  Fermion operator itself. A basis of the LLL wave
functions in the Landau gauge is given by
$$\{ \phi_n(z \bar{z})|n\in Z \} $$
where
\begin{eqnarray}
\phi_n ( z \bar{z}) & = & {1 \over \sqrt{L_x}} e ^{2\pi inx / L_x}
\left( {B \over \pi \hbar} \right)^{1 / 4}
\left( {2\pi \hbar  \over B} \right)^{1 / 2}
e ^{-(B / 2\hbar) (y-2\pi n/BL_x)^2 }
\nonumber \\
& = & \phi_n (\bar{z})  e^{-By^2 / 2 \hbar }.
\label{Landaugauge}
\end{eqnarray}
They are normalized with the measure $d^2z/\pi=(B/2\pi\hbar)dxdy$.
Fermion operator restricted on the LLL is expanded by $\phi_n(z {\bar z})$;
$\Psi_0(z {\bar z})=\sum c_n \phi_n(z {\bar z})$. These $c_n$'s are
anti-commuting
operators. From eq.(\ref{cylinder1}) the 1D representation of $\phi_n(z {\bar
z})$
is given by
\begin{equation}
\langle s | n \rangle
=\left( {{B \over {4\pi \hbar }}} \right)^{1/ 4}e ^{-\left( {\hbar / 2B}
\right)
\partial_s^2}\phi_n\left( {\bar z=\sqrt {B/ 2\hbar } \ s} \right)
={1 \over {\sqrt {L_x}}}e ^{i2\pi ns/ L_x}.
\label{1Dwavefunction}
\end{equation}
Therefore the 1D representation of the LLL Fermion operator $\Psi(s)=\sum c_n
\langle s|n \rangle$
satisfies the usual 1D Fermion anti-commutation relation,
$\{ \Psi(s), \ \Psi^{\dagger} (s') \} =\delta (s-s')$. From
(\ref{Landaugauge}),
Fermion operator in LLL in Landau gauge $\Psi_0(z \bar{z})$ is shown to
be written by its 1D representation $\Psi(x)$ as
\begin{equation}
\Psi_0(z, \bar{z})=\sqrt{{2\pi\hbar \over B}} \left( {B \over \pi \hbar}
\right)^{1/4}
\exp \left( -{B \over 2\hbar} \left( y+{i\hbar \over B} \partial_x \right)^2
\right)
\Psi(x).
\end{equation}
Then an equal-time correlator of the 2D Fermion operator becomes
\begin{eqnarray}
G(x,y;x',y') &=& \langle \Psi_0^{\dagger} (z, \bar{z}) \Psi_0(z',\bar{z'})
\rangle
\nonumber \\
&=& \sqrt{4\pi \hbar \over B} e^{-(B/2\hbar)[(y-i\hbar\partial_x/B)^2
               +(y'+i\hbar\partial_x'/B)^2]}
 \langle \Psi^{\dagger} (x) \Psi(x') \rangle.
\end{eqnarray}
If we consider a translational invariant system under $x$,
$\langle \Psi^{\dagger} (x) \Psi(x') \rangle$ depends only on its relative
distance
$r=x-x'$ and the above expression for $G$ is simplified as
\begin{equation}
G(x,y;x',y')=
\sqrt{4\pi \hbar \over B} e^{-(B/2\hbar)[(y-i\hbar\partial_r/B)^2
+(y'-i\hbar\partial_r/B)^2]}
 \langle \Psi^{\dagger} (r) \Psi(0) \rangle
\label{fermion-corr}
\end{equation}
All the information of a 2D correlator in the LLL (or on phase space) is
contained in its 1D correlator. This expression suggests that,
if a 1D Fermion system behaves as a Luttinger liquid,  not only the
edge but also the bulk correlator in the LLL will be characterized by its
Luttinger exponent.
\par
Here we consider the simplest example, $\nu =1$ (integer quantum Hall) droplet.
Consider a state
$ |G \rangle= \sum_{i=N_1}^{N_2-1} c_i^{\dagger}|0 \rangle.$
There are two Fermi surfaces with momenta $p_F^{(i)}=2\pi\hbar N_i/L_x$
\ ($i=1,\ 2$).  1D correlation function of Fermion operator for this state is
given by
\begin{eqnarray}
\langle \Psi^{\dagger} (r) \Psi(0) \rangle
 &=&  \sum_{i=N_1}^{N_2-1} {1 \over L_x} e^{-i2\pi n r/L_x}
  =   {1 \over L_x} {e^{-ip_F^{(1)} r/\hbar}-e^{-ip_F^{(2)} r/\hbar}  \over
      1-e^{-i2\pi r/L_x}.  \nonumber \\
  &\mathrel{\mathop{\kern0pt\longrightarrow}\limits_{L_x\to \infty }}&
{-i \over 2 \pi r} (e^{-ip_F^{(1)} r/\hbar}-e^{-ip_F^{(2)} r/\hbar}).
\end{eqnarray}
At the edge $y=y'=y_F^{(1)}=p_F^{(1)}/B$, 2d correlator for this state
becomes (from eq.(\ref{fermion-corr}))
\begin{eqnarray}
G(x,y_F^{(1)};x',y_F^{(1)}) & \approx &
     -i \sqrt{{\hbar \over B \pi}}
      e^{-ip_F^{(1)}r/\hbar} e^{\hbar \partial_r^2/B} {1 \over r}
     \nonumber \\
&=& -{i \over 2 \pi} e^{-ip_F^{(1)}r/\hbar} \int e^{-B (r-r')^2/4 \hbar}
    {1  \over r'} dr'.
\end{eqnarray}
This behaves $1/r$ for long-distance $r \gg l=(\hbar /B)^{1/2}$.
On the other hand the commutator damps exponentially if $y$ and $y'$
are  far from the edges. To see this let $p_F^{(1)}$ be $-\infty$
and $p_F^{(2)}$ be $+\infty$. Then
$\langle \Psi^{\dagger}(r) \Psi(0) \rangle =\delta (r)$ and accordingly 2D
correlator damps as
\begin{equation}
G(x,y;x',y')=e^{-(B/\hbar)[(y-y')^2+r^2]} e^{iBr(y+y')/2\hbar}.
\end{equation}
More generally eq.(\ref{fermion-corr}) shows that if $y$ and $y'$ are not on
the edges the exponential-damping factors don't vanish and the correlator damps
exponentially.
\par
In the fractional quantum Hall state it has been discussed \cite{wen}
\cite{stonefisher}
that 2D correlator at the edge has a power law  behaviour
\begin{equation}
G(x,y_F^{(1)};x',y_F^{(1)}) \sim   {e^{-ip_F^{(1)}r/\hbar} \over r^m}
\end{equation}
for $\nu=1/m$. This indicates that the 1D correlator for the Laughlin state
should also
behave in the same power law. On the other hand it is known \cite{kawakamiyang}
that Fermion correlator of the Calogero-Sutherland model has a different power
behaviour $r^{-\xi}$ whose exponent is $\xi=(m+1/m)/2$. In order to obtain
the correct power corresponding to the Laughlin state, we must decouple the
left-moving sector and the right-moving sector in the Calogero-Sutherland model
by the chiral constraint \cite{kawakami} or by redefining a decoupled Fermion
operator
\cite{stonefisher}. These procedures are necessary because in the quantum Hall
effect
the edge states at a different edges don't interact strongly but in the
Calogero-Sutherland
model they are scattered each other and mixed. In the narrow channel limit
where the Laughlin state becomes equivalent to the ground state of the
Calogero-Sutherland
model  the edge states at different edges become to interact and the power of
the
Fermion correlator will be changed. We would like to discuss it in a separate
paper.
\par
Next let us consider a 2D density operator in the LLL.
Projection of the density operator on the LLL has been discussed by
Girvin et.al.
\cite{girvin} to study the bulk collective excitations in the FQHE.
It also played an important role in $W_{1+\infty}$ symmetry
of the quantum Hall effect \cite{w}. Girvin et. al. considered
the projection by using an explicit form of the symmetric gauge wave functions.
Here we will use a gauge-independent formalism. 2D density operator
$\rho({\bf k})=\sum_i e^{-i{\bf k}\cdot{\bf x_i}}$ contains matrix
elements to mix different Landau level wave functions. This is easily
seen by rewriting $\rho({\bf k})$ in terms of $(\Pi_x, \Pi_y)$ and
guiding coordinates $(X,Y)$;
\begin{eqnarray}
\rho({\bf k}) &=& \sum_i \exp \left (-ik_x \left( X+{\Pi_y \over B} \right)_i
-ik_y \left( Y-{\Pi_x \over B} \right)_i \right)
\nonumber \\
&=& e^{-{\bf k}^2/4B} \sum_i e^{-k_z a_i^{\dagger}} e^{-i{\bf k}\cdot{\bf X_i}}
e^{k_{\bar{z}}a_i},
\end{eqnarray}
where $k_z=(k_x-i k_y)/\sqrt{2B}$.
The creation operator $a_i^{\dagger}=(\Pi_x+i \Pi_y)_i/\sqrt{2B\hbar}$
excites  LLL wave functions to a higher level.
Noting that a LLL state $|LLL \rangle$ satisfies $a_i|LLL \rangle=\langle
LLL|a_i^{\dagger}=0$,
projection of $\rho({\bf k})$ on LLL is given by
\begin{equation}
\rho_0({\bf k})=P_0 \rho({\bf k}) P_0=e^{-{\bf k}^2/4B} \sum_i e^{-i{\bf
k}\cdot{\bf X_i}}
\equiv e^{-{\bf k}^2/4B} W({\bf k}).
\end{equation}
$P_0$ is a projection operator onto the LLL.
\par
In the second quantized form,
 $W({\bf k})$ and  $\rho({\bf k})$
are written as
\begin{eqnarray}
W({\bf k}) &=& \int \Psi^{\dagger}(s) e^{-i(k_x X+k_y Y)} \Psi(s) ds
\nonumber \\
&=& \int \Psi^{\dagger}(s) e^{-i k_y Y/2}e^{-i k_x X} e^{-i k_y Y/2}  \Psi(s)
ds
\nonumber \\
&=&   \int \Psi^{\dagger}(s+ {\hbar k_y \over 2B}) \Psi(s-{\hbar k_y \over 2B})
e^{-ik_x s} ds  \label{w(k)} \\
\rho_0({\bf k}) &=&  \int e^{-{\bf k}^2/4B}
 \Psi^{\dagger}(s+{\hbar k_y \over 2B}) \Psi (s-{\hbar k_y \over 2B})
e^{-ik_x s} ds.  \label{rho(k)}
\end{eqnarray}
Here we used $X|s \rangle=s|s \rangle$ and $Y|s \rangle=p_s/B|s \rangle$.
These expressions are also given in \cite{wadia} in a different context.
\par
In the coordinate space, they become
\begin{eqnarray}
W({\bf x}) &=& \int W({\bf k}) e^{i {\bf k} \cdot {\bf x}}
  {d^2 {\bf k} \over (2\pi)^2 } \nonumber \\
&=&  \int \Psi^{\dagger}(x+{\hbar k_y \over 2B}) \Psi(x-{\hbar k_y \over 2B})
e^{ik_y y} {dk_y \over 2 \pi}
\nonumber \\
\rho_0({\bf x}) &=&  \int \rho_0({\bf k}) e^{i {\bf k} \cdot {\bf x}}
  {d^2 {\bf k} \over (2\pi)^2 }=e^{\hbar {\bf \nabla}^2/4B} W({\bf x})
\nonumber \\
&=& {B \over\pi \hbar} \int e^{-B ({\bf x}-{\bf x'})^2/\hbar} W({\bf x'})
d^2 {\bf x'}.
\end{eqnarray}
Strictly speaking,
these operators must be normal ordered and it gives a central extension term to
commutators of $W({\bf x})$ and $\rho({\bf x})$.
\par
$W({\bf x})$ is the Wigner distribution function of 1D Fermions and
it is consistent with the fact that the 2D space of the LLL is a phase space
for its corresponding 1D Fermions.
$\rho_0({\bf x}) $ is called the
Husimi distribution function, which is a coarse-grained
distribution of $W({\bf x})$. The Wigner distribution
 has a negative value
 while the Husimi distribution is everywhere positive definite.
If $W({\bf x})$ is spatially constant within the magnetic length
$l=\sqrt{\hbar/B}$,
these two distributions have the same behaviour.
\par
Integrating the 2D density operator, we obtain
\begin{eqnarray}
Q(x) & \equiv & \int_{-\infty}^{\infty} \rho_0({\bf x}) dy   \nonumber \\
&=& e^{\hbar \partial_x^2/4B} \int \Psi^\dagger(x+{\hbar k_y \over 2B})
\Psi(x-{\hbar k_y \over 2B}) e^{- \hbar k_y^2/4B} e^{i k_y y} {dk_y \over 2\pi}
dy
\nonumber \\
& = & e^{\hbar \partial_x^2/4B} \rho(x)
=\sqrt{{B \over \pi \hbar}} \int_{-\infty}^{\infty} e^{-B(x-x')^2/\hbar}
\rho(x') dx'
\end{eqnarray}
where $\rho(x)=\Psi^\dagger(x) \Psi(x)$ is the 1D density operator and
we call $Q(x)$ a charge operator.
\par
Expectation value of the 2D density operator for a translational invariant
state under $x$ is given by
\begin{eqnarray}
\langle \rho_0({\bf x}) \rangle &=& e^{\hbar \partial_x^2/4B}
    \int \langle \Psi^\dagger(x+{\hbar k_y \over 2B}) \Psi(x-{\hbar k_y \over
2B})
          \rangle
     e^{- \hbar k_y^2/4B} e^{i k_y y} {dk_y \over 2\pi} \nonumber \\
& = & e^{\hbar \partial_x^2/4B}
\int  \langle \Psi^\dagger(\hbar k_y/2B) \Psi(-\hbar k_y/2B) \rangle
     e^{- \hbar k_y^2/4B} e^{i k_y y} {dk_y \over 2\pi} \nonumber \\
&=& {B \over 2 \pi \hbar} e^{B \partial_k^2/4\hbar} n(k) =
   \sqrt{{B \over \pi \hbar}} \int e^{-\hbar (k-k')^2/B} n(k') {dk' \over 2
\pi}
\end{eqnarray}
where $n(k)$ is the momentum distribution of the 1D system
\begin{equation}
n(k)=\int \langle \Psi^{\dagger}(s/2) \Psi(-s/2) \rangle e^{iks} ds.
\end{equation}
Similarly a correlation function of the charge operator $Q(x)$ for a
translational
invariant state under $x$ is  written in terms of the 1D correlators;
\begin{equation}
\langle Q(x) Q(x') \rangle =e^{\partial_r^2/2B} g(r)
=\sqrt{{B \over 2 \pi \hbar}} \int e^{-B(r-r')^2/2 \hbar} g(r') dr'
\end{equation}
where $g(r)=\langle \rho(r) \rho(0) \rangle$.
In the narrow channel limit of the FQH state, the Laughlin state coincides with
the ground state of the Calogero-Sutherland model as shown in the
previous section and  the
1D density correlation of the  Laughlin state
has  a power law behaviour with a oscillation term $\cos (2k_F r)$.
Since the narrow channel condition eq.(\ref{narrow}) means that $k_F \ll
l^{-1}$
and the oscillation $\cos (2k_F r)$ is slow compared with the magnetic length,
the correlation function of the charge operator $Q(x)$  becomes same
as  $g(r)$ for $|x-x'| \gg l$.

\section{Holomorphic Representation of 1D Wave Functions}
So far we have considered the 1D representation of the LLL wave functions.
In section 2, it is shown that a sufficient condition for
equivalence of the Laughlin state and the ground state of the
Calogero-Sutherland
model is that the width of the cylinder droplet is narrow compared
with the magnetic length. It has been pointed out by many people
\cite{wen,stonefisher,kawakami}, however,
that the edge state of a non-narrow droplet of FQH state
behaves as the chiral Tomonaga-Luttinger liquid whose exponent is the same
as the chiral constrained Calogero-Sutherland model \cite{kawakami}. This
suggests that
there is more deeper relation between the Laughlin state and
the ground state of the Calogero-Sutherland model. In this section we
use the holomorphic representation of 1D Fermions to search for  universal
properties common in the Laughlin state and the Calogero-Sutherland model
\cite{brink}.
\par
Holomorphic representation of a 1D system whose coordinate is $s$
is constrained in terms of the coherent state
\begin{equation}
 a|\bar{z} \rangle =\bar{z} |\bar{z} \rangle,
\end{equation}
where a is an annihilation operator
\begin{equation}
a={1 \over \sqrt{2 \alpha}} (s + \alpha \partial_s), \ \ \
[a, a^{\dagger}]=1.
\end{equation}
$\alpha$ is an arbitrary constant. Coordinate representation of
$| \bar{z} \rangle$ is
\begin{equation}
\langle s| \bar{z} \rangle = c(z, \bar{z})
      e^{-s^2 /2 \alpha + \sqrt{2/ \alpha} s \bar{z} }.
\end{equation}
The normalization constant $c(z, \bar{z})$ is determined from
$\int |\langle s| \bar{z} \rangle |^2 ds =1$ as
\begin{equation}
|c(z, \bar{z})|^2= {1 \over \sqrt{\pi \alpha}}
    e^{-|z|^2-(z^2 +\bar{z}^2)/2}.
\end{equation}
If we write $z=(x+iy)/\sqrt{2 \alpha}$ and choose $c$ and $\alpha$ by
\begin{equation}
c(z, \bar{z})= \left( {1 \over \pi \alpha} \right)^{1/4}
e^{-z^2-y^2/2 \alpha}, \ \ \ \alpha={\hbar \over B},
\end{equation}
$\langle s| \bar{z} \rangle $ becomes equal to the 1D representation
of LLL wave functions in the Landau gauge, eq.(\ref{1Drep}).
Thus the 1D representation of LLL wave functions is an inverse-transformation
of the holomorphic representation of 1D wave functions.
A different choice of a phase of $c$ corresponds to a different choice
of the gauge in 2D LLL wave functions. By using this holomorphic
representation, we
can interpret 1D wave functions as wave functions in the LLL
in 2D. From eq.(\ref{cylinder1}), LLL representation of a
1D wave function $\phi(s)$ is given by
\begin{equation}
\Psi(\bar{z}) e^{-B y^2/2 \hbar}
\end{equation}
where
\begin{equation}
\Psi(\sqrt{B /2 \hbar} \ s)= \left( {4 \pi \hbar \over B} \right)^{1/4}
 e^{(\hbar / 2B ) \partial_s^2 } \phi(s).
\end{equation}
For the ground state of the Sutherland model (\ref{Sutherland}),
its LLL representation is
\begin{equation}
\Psi(\bar{z_1}, \cdot \cdot \bar{z_N}) =
e^{\sum_i (\partial_{\bar{z_i}})^2/4} \prod (\omega_i -\omega_j)^m,
\end{equation}
where $\omega_i=e^{i 2 \pi(x_i-iy_i)/L_x}=e^{i(2\pi/L_x)\sqrt{2\hbar/B}
\bar{z_i}}$
and we neglected a constant normalization factor.
For a large $L_x$ limit, $(\omega_i -\omega_j)$ can be approximated
by $(\bar{z_i}-\bar{z_j})$ and
\begin{equation}
\Psi(\bar{z_1}, \cdot \cdot \bar{z_N}) \approx
e^{\sum_i (\partial_{\bar{z_i}})^2/4} \prod (\bar{z_i}-\bar{z_j})^m.
\end{equation}
Wave functions in a magnetic field are generally determined by positions of
zeros.
For the Laughlin state, its zeros of some particle are located at
positions of the other particles with degeneracy $m$. The above wave
function also has $m$ zeros around positions of the other particle but not
exactly on them. This discrepancy is caused by the differential operator
$e^{\sum_i (\partial_{\bar{z_i}})^2/4}$.
To study the effect of this operator, we consider a two-particle case.
\begin{equation}
\Psi(\bar{z_1}, \bar{z_2}) = e^{\sum_i (\partial_{\bar{z_i}})^2/4}
(\bar{z_1}-\bar{z_2})^m.
\end{equation}
By defining $z=z_1-z_2$, the wave function becomes
\begin{equation}
\Psi(\bar{z})=e^{(\partial_{\bar{z}})^2/2} (\bar{z})^m
=(\bar{z})^m +m(m-1)/2  \ (\bar{z})^{m-2} + \cdot \cdot \cdot.
\end{equation}
The $m$-th zero's of this wave function are shifted from $z=0$.
%
%
%
 Note that
the scale is given by the magnetic length; $z=(x+iy)/\sqrt{2} l$.
The effect of the exponential operator is, therefore, to shift the
$m$ zero's of a wave function, so that
the $m$ zeros are distributed around $\bar{z}=0$ with a distance of
the order of the magnetic length.
The discrepancy from the Laughlin state reduces to zero in a small magnetic
length limit
(or large $B$ limit). This simple exercise suggests that the $m$ zeros
of the Sutherland ground state in the holomorphic representation are
distributed around the positions of the
other particles with an extent of the magnetic length.
As Halperin discussed before \cite{halperin}, Laughlin state is not an exact
ground state for the Coulomb interaction but the true ground state wave
function
of the fractional Hall effect is given by distributing $m$ zeros around
other particles with a distance of the
 magnetic length. Since the Sutherland ground state
satisfies the above criterion, it may exhibit the
fractional quantum Hall effect on its phase space
and therefore may be classified in the same universality class as the Laughlin
state. Long-distance universality of the Laughlin state is also discussed by
\cite{fradkin}.
(On the contrary usual 1D Fermion system
can be interpreted as a droplet of the integer quantum Hall state on its phase
space.)
The Laughlin state and the ground state of the Calogero-Sutherland model
are different in short-distance but their long-distance behaviour for
bulk quantities  will be the same.
\par
The discussion of this section is not restricted to the $1/r^2$ model on a
cylinder geometry
(Sutherland model). $1/r^2$ model on  different geometries
or with different confining potentials will also exhibit the  fractional
quantum
Hall effect on their phase space. Different geometries (or different confining
potentials)  correspond to  different shapes of the droplet in the 2D phase
space.
Recently Brezin and Zee discussed a universal behaviour of the inverse-square
model (in a context of matrix model) with various confining potentials
\cite{brezinzee}.
 It will be interesting to
interpret the universality in terms of the language of the 2D Fermions in
the lowest Landau level.
\section{Summary and Discussions}
In this paper we studied  universal properties of the Laughlin state and the
ground
state of the Calogero-Sutherland model. Beyond  apparent similarities such as
the Jastrow form of the wave functions,
fractional charge and statistics of the quasi-hole excitations and
 hierarchy extensions, these two different systems are much more deeply
related.
The seeming difference of the dimensionality is only the difference of
representation
of wave functions and if we use the same representation they have a similar
long-distance behaviour. In particular, if the Laughlin state is confined in a
narrow
cylinder geometry, it exactly coincides with the ground state of the Sutherland
model. We also give a recipe to rewrite 2D operators acting on states in the
lowest Landau level in terms of usual 1D Fermion operators.
\par
These two systems are expected to have the same long-distance behaviour
but there is one difference. In the fractional quantum Hall effect, edges
states
 at separate boundaries don't interact strongly while in the
Calogero-Sutherland
model they interact through $1/r^2$ interaction. This causes the difference
of the exponents of the Fermion correlators. The same correlation can
be obtained by decoupling left and right going sectors in the
Calogero-Sutherland
model through the chiral constraint \cite{kawakami} or through redefining a
Fermion operator \cite{stonefisher}.
\par
Here we comment on a possible new type of 1D liquid,
{\it fractional Tomonaga-Luttinger liquid}.
Usual 1D Fermions can be interpreted as a droplet of an integer Hall state
on its phase space and the low energy collective excitations are described
 by the edge states, density deformation of the boundary of the droplet.
The filled region in the droplet is thus completely filled and this causes
imcompressibility of the droplet. Current-current interaction of Fermions
only rearranges the density fluctuations along edges. In the Calogero-
Sutherland model, however, the bulk property is  changed more than a
rearrangement of edges. The filled region on the phase space becomes
fractionally
filled and can be interpreted as a droplet of the fractional quantum Hall
state.
The low energy collective excitations are again edge states.
Important properties of this liquid are the existence of fractionally charged
quasi-hole excitations with fractional statistics.
This new type of 1D liquid will give a new paradigm of 1D integrable models
as Haldane stresses in his recent papers \cite{haldane2}.
\par
Finally it is also worth noting that $1/r^2$ interaction in 1D is "statistical"
rather than "dynamical" \cite{poly,bernardwu}. In 2D Chern-Simons gauge field
plays the same role of statistical interaction
and the effective field theory of the quantum Hall effect
is described by the Chern-Simons gauge theory.  Then it can be expected that
an effective field theory of the Calogero-Sutherland model is also described
by using the Chern-Simons gauge on its phase space.
\newpage
\section*{Appendix: Proof of the last equality in (\ref{cylinder1})}
\setcounter{equation}{1}
\renewcommand{\theequation}{A.\arabic{equation}}
Set an operator A
\begin{equation}
A= e^{-Bs^2/2} e^{- \partial_s ^2 /4B} e^{Bs^2} \ 2^{s \partial_s}
\end{equation}
and a function $g_k(s)$
\begin{equation}
g_k(s)=A \ e^{iks}.
\end{equation}
It is easy to show that
\begin{equation}
sA=A(s+{\partial \over B}), \ \ \ \partial_s A=A \partial_s.
\end{equation}
Then
\begin{equation}
\partial_k g_k(s) = iAse^{iks} =i(s-\partial_s/B) A e^{iks}
                =   (is + {k \over B}) g_k(s)
\end{equation}
and $g_k(s)$ is written as
\begin{equation}
g_k(s)=e^{k^2/2B +iks} g_{k=0}(s).
\end{equation}
Next $g_0(s)$ satisfies
\begin{equation}
\partial_s g_o(s) = \partial_s A \cdot 1 =0
\end{equation}
and $g_0(s)$ is $s$-independent.
{}From the definition of $g_s(k)$, $g_0(0)$ is given by
\begin{eqnarray}
g_0(0) &=& (e^{-\partial_s^2/4B} e^{B s^2} )_{s=0}
      = \sum_{n=0}^{\infty} \left( -{B \over 4 B} \right)^n {1 \over (n!)^2}
     \partial_s^{2n} s^{2n}
    \nonumber \\
   &=& \sum_{n=0}^{\infty} \left( -{1 \over 4} \right)^n {(2n)! \over (n!)^2}
     =  {1 \over \sqrt{2} }
\end{eqnarray}
and therefore
\begin{equation}
g_k(s)= g_0(0) e^{k^2/2B +iks}=g_0(0) e^{-(\partial_s)^2 /2B} e^{iks}.
\end{equation}
Hence it is proved that
\begin{equation}
A=g_0(0) e^{-(\partial_s)^2 /2B}.
\end{equation}

\end{document}